\documentclass[10pt,journal,  oneside, twocolumn]{IEEEtran}

\usepackage{multirow}
\usepackage{latexsym}
\usepackage{graphicx}
\usepackage{float}
\usepackage{amsmath}
\usepackage{amsthm}
\usepackage{lipsum}
\usepackage{subfig}
\usepackage{graphicx}
\usepackage{authblk}
\usepackage{bm}
\usepackage{booktabs}
\usepackage{amsthm}
\usepackage[section]{placeins}
\usepackage{soul}

\usepackage{colortbl}

\makeatletter  
\newif\if@restonecol  
\makeatother

\usepackage[linesnumbered,ruled,vlined]{algorithm2e}
\usepackage{algpseudocode}  
\usepackage{amsmath}

\usepackage{amssymb}

\usepackage{mathrsfs}
\usepackage{subfig}
\usepackage{caption}
\newcounter{problem}
\newcommand{\theprob}{P\refstepcounter{problem}\theproblem}
\newcommand{\probref}[1]{P\ref{#1}}
\SetKwInput{KwInitialize}{Initialize}

\renewcommand{\baselinestretch}{1.0}

\begin{document}

\title{Multi-hop RIS-aided Learning Model Sharing for Urban Air Mobility}


\author{Kai Xiong~\IEEEmembership{Member,~IEEE}, Hanqing Yu, Supeng Leng,~\IEEEmembership{Member,~IEEE}, Chongwen Huang,~\IEEEmembership{Member,~IEEE}, Chau Yuen,~\IEEEmembership{Fellow,~IEEE} 

\thanks{

K. Xiong, H. Yu, and S. Leng are with School of Information and Communication Engineering, University of Electronic Science and Technology of China, Chengdu, 611731, China; and, Shenzhen Institute for Advanced Study, University of Electronic Science and Technology of China, Shenzhen, 518110, China.
}

\thanks{
C. Huang is with College of Information Science and Electronic Engineering, Zhejiang University, Hangzhou 310027, China.
}

\thanks{
C. Yuen is with School of Electrical and Electronics Engineering, Nanyang Technological University, 639798, Singapore. 
}







}


\maketitle



\begin{abstract}
Urban Air Mobility (UAM), powered by flying cars, is poised to revolutionize urban transportation by expanding vehicle travel from the ground to the air. 
This advancement promises to alleviate congestion and enable faster commutes.
However, the fast travel speeds mean vehicles will encounter vastly different environments during a single journey. As a result, onboard learning systems need access to extensive environmental data, leading to high costs in data collection and training. 
These demands conflict with the limited in-vehicle computing and battery resources.
Fortunately, learning model sharing offers a solution. Well-trained local Deep Learning (DL) models can be shared with other vehicles, reducing the need for redundant data collection and training. However, this sharing process relies heavily on efficient vehicular communications in UAM. 
To address these challenges, this paper leverages the multi-hop Reconfigurable Intelligent Surface (RIS) technology to improve DL model sharing between distant flying cars. 
We also employ knowledge distillation to reduce the size of the shared DL models and enable efficient integration of non-identical models at the receiver. 
Our approach enhances model sharing and onboard learning performance for cars entering new environments. Simulation results show that our scheme improves the total reward by $85\%$ compared to benchmark methods.





\end{abstract}


\IEEEpeerreviewmaketitle

\section{Introduction}

{
\IEEEPARstart {A}{s} urbanization and population growth continue, the road traffic in central cities worldwide has become overloaded.
To relieve traffic congestion and infrastructure expenses, automakers are exploring flying cars, which expand transportation into Near-Ground Space (NGS) \cite{{9345783Gaofeng}}.
Without the constraints of road infrastructure, flying cars promise faster commutes and reduced costs associated with road construction and maintenance. Urban Air Mobility (UAM) is the transportation paradigm that emerges from this development, facilitating the movement of people and goods via near-ground airspace. According to a report from Morgan Stanley \cite{Morgan2019}, the UAM industry could reach a value of \$1.5 trillion by 2040, comparable to the potential market size for autonomous driving.

While flying cars offer the promise of efficient transportation, they also introduce significant challenges to aviation safety. The most notable challenge is the need for complex 3D environmental detection, which is essential for autonomous flying. Flying cars must accurately detect both aerial (e.g., other flying cars, drones, birds) and ground entities (e.g., pedestrians, ground vehicles) \cite{9447255Adam}. Fortunately, point cloud data generated by onboard LiDAR systems helps meet this challenge. Point clouds provide detailed geometric information about the surroundings, allowing for accurate 3-Dimensional (3D) object detection \cite{Tosteberg2017SemanticSO}.

However, flying cars encounter various environments during a single journey, making it impractical to collect and process all relevant data. A feasible alternative is for vehicles to acquire well-trained DL models from local cars in the area. These models can either be used directly for detection or further retrained. However, sharing DL models between flying cars is challenging because of the large safe separation for fast aviation as well as obstacles, such as skyscrapers, that can block signal transmission \cite{{10149027Xiaosha}}.

Reconfigurable Intelligent Surfaces (RIS) offer a solution to these challenges. RIS is a transformative 6 Generation (6G) technology that manipulates radio waves to customize signal propagation. Each RIS element can combine and reflect signals in a specific direction, which enables better signal transmission between distant flying cars, even in environments with non-line-of-sight (NLoS) conditions \cite{9500188Xu}. However, despite the promise of RIS technology, several difficulties remain for DL model sharing in UAM.

First, the onboard DL models of flying cars are often complex and contain numerous parameters, making them difficult to transmit over wireless networks. Second, different flying cars may embrace different DL model structures. For instance, flying cars in urban environments may require more complex models than those in rural areas, where there are fewer obstacles and entities to detect. These differences make it hard to directly integrate DL models in a federated learning fashion, which requires identical parameter structures \cite{9831009Beibei}.

To overcome these difficulties, we exploit knowledge distillation (KD) to compress DL models for transmission and facilitate the integration of non-identical models. KD allows a larger, well-trained model (the teacher) to transfer its knowledge to a smaller, lightweight model (the student) \cite{2020Empowering}. This process reduces the size of the shared DL model, enabling efficient transmission, and allows for non-identical model integration without requiring the same structure for all models.

Our proposed DL model sharing scheme includes three key components: model compression, propagation, and integration. 
KD is used to generate a lightweight model for transmission. 
Multi-hop RIS technology is employed to transmit the compressed model between flying cars. 
Finally, the received models are integrated into the onboard DL model of the receiver using KD method. Our scheme significantly improves onboard learning performance for cars entering new environments. The main contributions of this paper are as follows:

\begin{itemize}
\item 
We design a multi-hop RIS-aided DL model sharing scheme. This scheme improves long-distance communication between flying cars by reflecting signals through multi-hop RIS communication, overcoming obstacles like buildings. We also develop two phase shift optimization algorithms for RIS. One algorithm uses the Semi-Definite Relaxation (SDR) method to achieve optimal phase shifts, while the other offers a fast suboptimal solution with lower computational complexity. These options balance performance and time consumption, making the scheme adaptable for various UAM scenarios.



\item We propose a KD-based DL model compression and integration scheme. The local DL model is compressed into a lightweight version for transmission, significantly reducing overhead without sacrificing much performance. We design three types of propagated DL models (lightweight version) to meet different performance and transmission delay requirements. Additionally, we develop a multiple-teacher, single-student KD framework to facilitate the integration of diverse models at the receiver.


\item We formulate a joint optimization problem that minimizes both transmission delay and final learning performance after DL model sharing. To solve this, we propose a Block Coordinate Descent (BCD) method that iteratively optimizes the KD-based model selection and multi-hop RIS phase shift control. This approach balances transmission performance and DL model integration quality in resource-constrained systems.
Simulations illustrate that our proposed algorithms outperform the benchmarks in terms of transmission rate, model sharing performance, and optimization time consumption.

%
\end{itemize}

The remainder of this paper is organized as follows. Section II reviews related works. Section III presents the system model. Section IV proposes our solution for optimizing model propagation and integration. Section V provides simulation results and performance analysis. Finally, Section VI concludes the paper.

\section{Related Work}
As engineers and researchers develop flying cars for urban transportation, 3D environmental perception becomes crucial for safe aviation. Sharing DL models between vehicles can help flying cars quickly adapt to new environments by leveraging the knowledge from local cars. However, this requires efficient long-distance communication.
Recent work has explored multi-hop RIS as a promising technology for enabling non-line-of-sight (NLoS) communication in dense urban environments. Additionally, knowledge distillation has emerged as a popular technique for compressing and transferring DL models. Below, we review the literature on flying cars, RIS communication, and knowledge distillation, respectively.
}

 
\subsection{Flying Car}
Flying cars, capable of both road driving and vertical takeoff and landing (VTOL) flight, are being developed by several companies \cite{7420801Kaushik}. For example, EHang launched its autonomous aerial vehicle, the “EHang 184,” in 2016. The operating costs of flying cars are competitive with traditional taxis, suggesting they could become a viable commercial transport option. 
Leading automobile companies like Audi, Toyota, Geely, and Xiaopeng are preparing to launch commercial aircraft in 2024 \cite{Ehang3248}.
Additionally, multiple flying cars can also leverage distributed sensing to improve safe aviation by sharing DL models \cite{7452570Fang}. 
It confirms the available commercial model of flying cars, with an expected payback on public transport.
These manufacturing, transportation, and related services involved in flying cars are becoming crucial economic growth points for regional industrial integration and employment.
However, communication between flying cars remains challenging, especially due to Vehicle-to-Vehicle (V2V) link blocking in urban areas \cite{Atrianfar}.


\begin{figure*}
\centering
     \includegraphics[width=.8\textwidth]{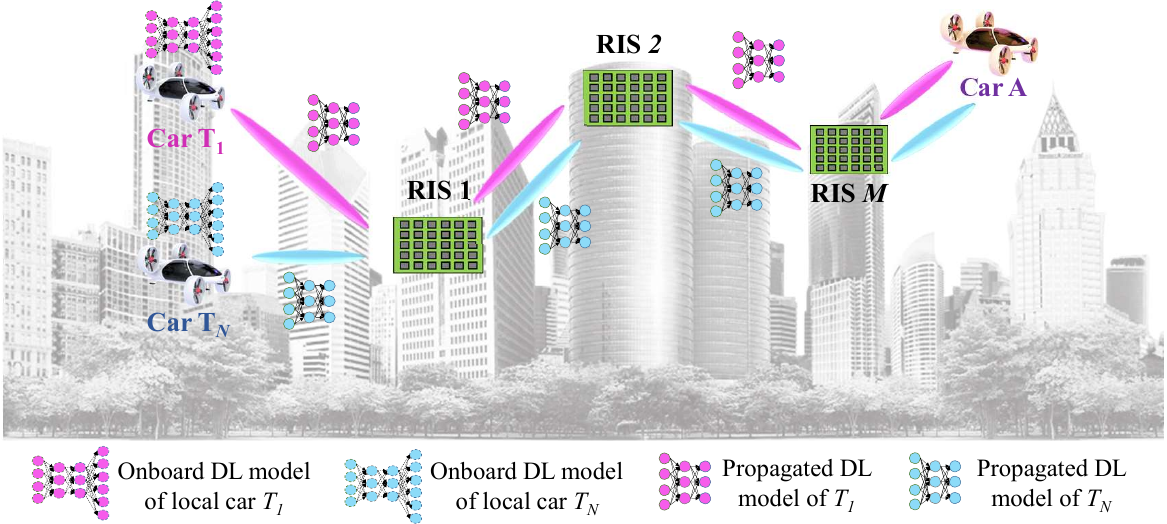} 
     \caption{Multi-hop RIS-aided DL model sharing in UAM.} 
\label{total_scenario}
\end{figure*}

\subsection{RIS Communication}
RIS is a meta-surface that can manipulate incoming electromagnetic waves to reflect them in a desired direction \cite{9110869Chongwen}. The large-scale deployment of RISs can improve communication coverage \cite{9174910Kishk}. 
The authors in \cite{9410457Chongwen} showed the superiority of multi-hop RIS communication over a single-RIS. In addition, they jointly optimized the transmission beamforming of the base station-RIS, RIS-RIS, and RIS-user links through the DDPG method.
Moreover, the work in \cite{9829192Xiaoyan} pointed out that multi-hop RIS systems can provide higher channel diversity and bypass obstacles, ensuring reliable communication even in dense urban areas. 
Although previous literature thoroughly researched the multi-hop RIS capability, it remains largely unexplored the multi-hop RIS communication optimization for specific applications, like DL model propagation.
Besides, this paper first leverages the multi-hop RIS communication to provision efficient collaborative learning between flying cars.

{
\subsection{Multi-hop Communication}
A single agent may not always have access to all the information it requires. Acquiring long-range information can help agents avoid short-sighted decisions and enhance the training of the onboard DL model. As a result, effective multi-hop communication is essential, given the limitations of the realistic communication range \cite{Wang3545946_3598667}. Nguyen \textit{et al.} \cite{Nguyen9924192} introduced a two-hop communication protocol that integrates a dynamic resource allocation strategy to maximize the number of clients participating in federated learning (FL). Their simulations confirmed that a multi-hop network can improve FL accuracy. Zhang \textit{et al.} \cite{10345598Zhang} proposed a novel recursive training method for the encoder and decoder in multi-hop semantic communication systems. This method enables the system to handle distorted received messages and achieve higher transmission rates. Park \textit{et al.} \cite{Park9623651} developed low-complexity schemes that are provably efficient and achieve logarithmic regret growth in transmission learning, optimizing multi-hop throughput performance.

However, the aforementioned works primarily focus on improving multi-hop communication performance and often overlook the utilization of multi-hop information for training DL models. To address this gap, Pinyarash \textit{et al.} \cite{Pinyarash9154266} proposed a modular wireless edge system for FL, incorporating programmable multi-hop network control to ensure FL convergence and accuracy. Their experimental results demonstrated the potential of these algorithms to significantly accelerate FL convergence in wireless multi-hop networks. That said, this work did not consider the learning integration of non-identical DL model structures.

In contrast, our paper explores the application of the latest 6G technology, Reconfigurable Intelligent Surfaces (RIS), for DL model propagation using multi-hop RIS communication. Furthermore, we optimize both transmission and learning performance within the system, rather than focusing solely on communication optimization for learning purposes

}

\subsection{Knowledge Distillation}
This paper applies Knowledge Distillation technology to transfer well-trained knowledge from local cars to foreign cars.
Hinton \textit{et. al} \cite{2015HintonKD} first proposed KD as a teacher-student paradigm widely employed for DL model compression. 
There are several widely used methods for model compression, including knowledge distillation, network pruning \cite{Han2015DeepCC}, quantization \cite{Jiaxiang7780890}, and binarization \cite{Rastegari32}. 
However, the KD approach can transfer the knowledge of non-target labels from the DL model to the lightweight model.
By contrast, the other compression methods do not have this non-label learning ability.
Bhardwaj \textit{et al.} \cite{2020Empowering} used the KD method on the Pointnet learning and generated a lightweight learning model analogous to the Pointnet. 
However, most of the previous knowledge transfer studies focus on transferring the knowledge from one model to another model \cite{{Bo9416178},{9127823Chai}}.
This paper utilizes the KD in the multiple teachers and students design to achieve the non-identical model integration.
It would adapt to various DL models with different sizes and performances in heterogeneous multi-agent systems.

Overall, while significant progress has been made in DL model compression and RIS communication, their combination for UAM remains underexplored. This paper proposes a multi-hop RIS-aided DL model sharing scheme that leverages KD to improve onboard learning performance in flying cars.


\section{System Architecture}
{
The goal of this paper is to develop an efficient multi-hop RIS-based DL model sharing scheme for UAM. Our approach includes three main components: DL model compression, propagation, and non-identical DL model integration. The key challenge is to enable efficient communication and knowledge sharing between flying cars encountering diverse environments.

Fig.~\ref{total_scenario} illustrates our proposed UAM scenario. Flying car $A$ routinely cruises around a city. Occasionally, it may need to travel to a new region. Since car $A$ is unfamiliar with this new environment, it struggles to recognize specific 3D objects, such as regional migratory birds or landmarks. Consequently, car $A$ requires retraining of its onboard DL model to adapt to the new environment. However, retraining based solely on local data is time-consuming and inefficient.

Fortunately, local cars $T1$ and $T2$ possess well-trained DL models for the region. These models can be shared with car $A$ to accelerate its adaptation process. First, the local cars compress their onboard DL models using knowledge distillation. This produces a lightweight propagated version of the onboard model, which can be transmitted to car $A$ via multi-hop RIS communication. Once car $A$ receives the propagated models, it extracts the propagated model knowledge into its own onboard DL model by a KD-based integration method.
}

\begin{figure*}
\centering
     \includegraphics[width=.8\textwidth]{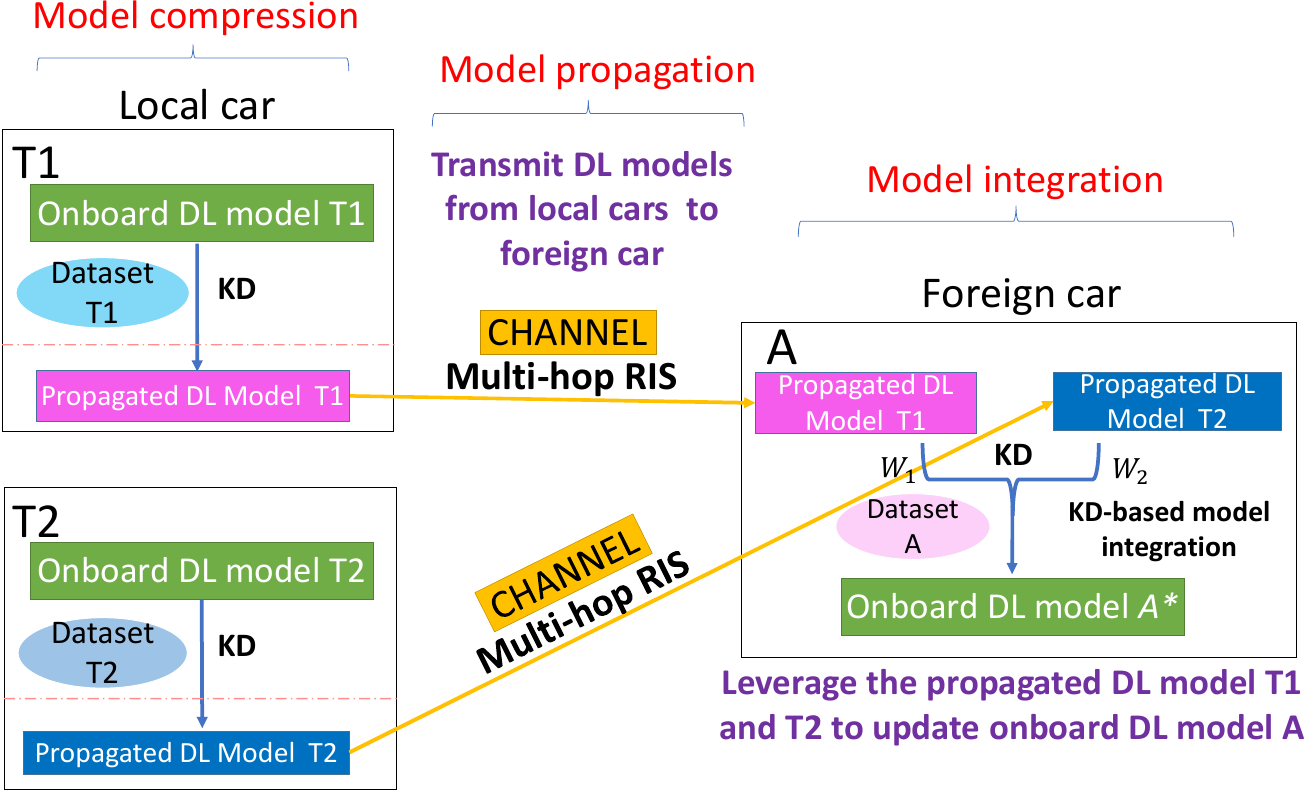} 
     \caption{DL model sharing architecture.} 
\label{total_architecture}
\end{figure*}

Fig~\ref{total_architecture} presents the proposed DL model sharing architecture.
In the DL model compression part, we leverage the KD approach to generate the propagated DL model.
The multi-hop RIS is engaged to transmit the propagated DL model significantly facilitating the coverage range of DL model propagation.
Finally, the knowledge is distilled from the propagated DL models to the onboard DL model of the foreign car in the DL model integration part. 
The proposed DL model sharing architecture can efficiently distill and propagate the knowledge from multiple local cars and boost the onboard DL model update of the car when it first arrives in a new region.
Hereafter, we separately characterize each component of the architecture in detail.


{
\subsection{KD-based DL Model Compression}
For efficient transmission, the shared DL model needs to be lightweight. To achieve this, we use knowledge distillation (KD) to compress the onboard DL model into a lightweight version, which we call the propagated DL model.

Accordingly, we can see that the KD framework involves two roles: a teacher (the well-trained, onboard DL model) and a student (the lightweight, propagated DL model). This paper applies the Pointnet neural network for onboard 3D point cloud detection. As shown in Fig.~\ref{BK_Structure}, Pointnet consists of four main components: input transform, feature transform, shared multi-layer perception (MLP), and final softmax output. The output is the probability distribution over object categories.

The investigated point cloud data is composed of a set of 3-D coordinates $\{\xi^{j}_i = (x_i, y_i, z_i) | i=1,2,..,n\}$, where $\xi^{j}_i$ is the $i$th point of the entity $j$.
$n$ is the number of points to depict an entity.
Point cloud learning provides the classification of the point cloud data $\xi^{j}$, which maps point cloud features to probability distributions over each category,
\begin{equation}
\begin{split}
\begin{aligned}
D_{pc}(\xi^{j}_i, \xi^{j}_2,\dots, \xi^{j}_n)\rightarrow K,
\end{aligned}
\end{split}
\label{sjdkfsdlkjfhasduifew}
\end{equation} 

\noindent where $K$ is the category of the point cloud object.
The output $K^j_{dist}$ is the category probability distribution of object $j$:
\begin{equation}
\begin{split}
\begin{aligned}
K^j_{dist} = \frac{exp(z_{jk})}{\sum_{k\in K} exp(z_{jk})},
\end{aligned}
\end{split}
\label{f7eryf8u32jhkjlsda}
\end{equation} 

\noindent in which $z_{jk}$ is the final logits vector. 
However, Pointnet has a large storage size of $6.2$ MB, making it inefficient for transmission. To reduce overhead, we design a propagated DL model with a simplified structure. The propagated DL model consists only of a shared MLP layer, reducing its size to $2.3$ MB.

\begin{figure}[h]
\centering
\includegraphics[width=.49\textwidth]{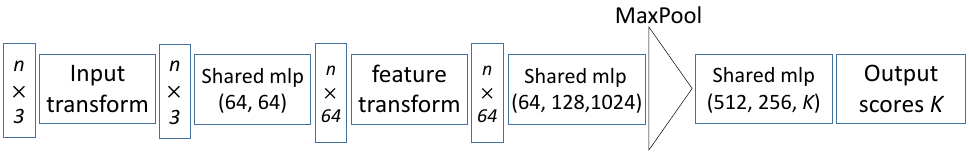} 
     \caption{Onboard DL model of car $T1$, i.e., Pointnet} 
\label{BK_Structure}
\end{figure} 

\begin{figure}[h]
\centering
     \includegraphics[width=.45\textwidth]{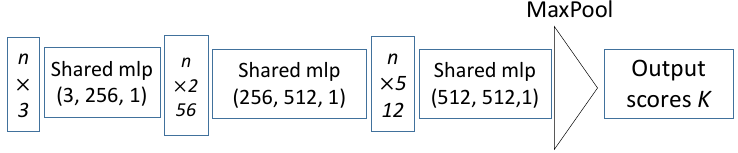} 
     \caption{Propagated DL model of car \textit{T1}.} 
\label{T1_structure}
\end{figure}
 
The KD process transfers the knowledge from the teacher (onboard DL model, i.e., Pointnet) to the student (propagated DL model). 
The loss function for KD training consists of two parts: a cross-entropy loss between the predicted vector $\bm{Z^{r}}$ and the true label $\bm{y}$, and an alignment loss between the student’s and teacher’s predicted logits, $Z^{r}$ and $Z^{OB_r}$, respectively. 
The overall KD loss function weightily combines the two parts, as given by \cite{2015HintonKD}:
\begin{equation}
\begin{split}
\begin{aligned}
min_{Z} \ L(Z) =\sum_{i=1}^n (1-\alpha)\mathscr{L}(\sigma(\bm{Z^{r}}(\bm{x}_i)), \bm{y}_i) \\
+ \alpha\tau^2 \mathscr{L}(\sigma(Z^{OB_r}(\bm{x}_i),\tau), \sigma(Z^{r}(\bm{x}_i), \tau)),
\end{aligned}
\end{split}
\label{dsfjsdhkgeriughprevnkl}
\end{equation} 

\noindent where $\tau$ represents the distillation temperature. 
$n$ is the number of points of the point cloud data. 
$\alpha$ is a weight parameter.
$\bm{x}$ is the training point cloud dataset, $\bm{y}$ represents the corresponding truth one-hot label, and $\mathscr{L}$ is the cross entropy to measure the gap between the teacher and student logits.
$\sigma$ indicates the softmax operation that converts the logits $Z$ to the probability outputs.
In addition, $\sigma(z,\tau)$ represents the temperature-augmented softmax operation. i.e.,
\begin{equation}
\begin{split}
\begin{aligned}
\sigma(\bm{z}_i,\tau) = \frac{exp(\bm{z}_i/\tau)}{\sum_j exp(\bm{z}_j/\tau)}.
\end{aligned}
\end{split}
\label{ktototototototv}
\end{equation}

After the KD training, the flying car acquires the propagated DL model with light storage size and acceptable detection accuracy.
The KD process produces a propagated DL model that is roughly $16\%$ the size of the original model, significantly reducing the overhead for transmission.
}

\subsection{Multi-hop RIS-aided DL Model Propagation}

We propose using multi-hop RIS communication to transmit the compressed DL models between flying cars. In dense urban environments, direct communication between flying cars may be blocked by skyscrapers or other obstacles. As shown in Fig.~\ref{total_scenario}, multi-hop RIS elements placed on building facades can reflect signals and bypass obstacles, improving signal coverage and overcoming non-line-of-sight (NLoS) issues.

The investigated multi-hop RIS system consists of $M$ number of local cars, i.e., $T1$, $T2$, \ldots, $T{M}$, one foreign car $A$, and $N$ number of RISs. 
We assume that each car possesses a single antenna for vehicle-to-vehicle (V2V) communication.
Moreover, the V2V communication MAC layer is supposed to employ the Time Division Multiple Access (TDMA) protocol. This indicates that there is no interference between each communication stream.

{
This V2V communication has three sub-links: the car-to-RIS link, the RIS-to-RIS link, and the RIS-to-car link. The received signal at car $A$.
Denote the channel matrices of the local car-RIS link, RIS-RIS link, and RIS-foreign car link as $\boldsymbol{g}_{m} \in \mathbb{C}^{N_1 \times 1}$, $\mathbf{H}_{i} \in \mathbb{C}^{N_{i+1}\times N_i}$, and $\mathbf{H}_{N} \in \mathbb{C}^{1\times N_N}$, respectively. 
Wherein, $N_i$ is the number of reﬂecting elements of the $i$th RIS.
Further, this paper assumes the Racian fading in the channels, thus $\mathbf{H}_i$ is given as,
\begin{equation}
    \mathbf{H}_i = \sqrt{\frac{\beta_H}{\beta_H + 1}}\bar{\mathbf{H}}_i + \sqrt{\frac{1}{\beta_H + 1}}\tilde{\mathbf{H}}_i,
\end{equation}
in which $\beta_H$ is the Racian factor, $\bar{H}_i$ is the the line-of-sight (LoS) component, and $\tilde{H}_i$ is the non-LoS component.
Similarly, the channel $\boldsymbol{g}_{m}$ follows Racian fading with Racian factor $\beta_g$.
Accordingly, the received signal of $Tm$ at car $A$ is,
\begin{equation}
    y_{m} = \Big(\prod_{i=1}^{N} \mathbf{\Theta}_{i} \mathbf{H}_{i}\Big) \boldsymbol{g}_{m} x_m + n_A
\end{equation}

\noindent where denote by $x_m$ the propagated DL parameters of car $Tm$. Denote by $\mathbf{\Theta}_{i} = \mathrm{diag}(e^{j\theta_{i,1}}, \ldots, e^{j\theta_{i,N_i}}) \in \mathbb{C}^{N_{i}\times N_i}$ the phase shift matrix of RIS $i$, where $\theta_{i,j}$ is the phase shift induced by $j$th element of the $i$th RIS.
$n_A$ is additive white Gaussian noise (AWGN) at the car $A$ with zero mean and variance $\sigma^2$.
It is assumed that there is no energy loss when the signal goes through RIS.
The $n$th element on the diagonal of the $\mathbf{\Theta}_{i}$ satisfy $|\mathbf{\Theta}_{i}(n, n)|=1$.
Then, we get the transmission rate of car $Tm$ to car $A$ as,
\begin{equation}
    R_m = B_m \log_2\left(1 + \frac{\left|\left(\prod_{i=1}^{N} \mathbf{\Theta}_{i} \mathbf{H}_{i}\right) \boldsymbol{g}_{m} x_m\right|^2}{\sigma_A^2}\right)
\label{sdafkhjaguyfgercbbaaosduf}
\end{equation}
where $\sigma_A^2$ is noise power and $B_m$ is the bandwidth.
We also assume the channel state information of these channels is perfectly known at both the cars and RISs. Although obtaining these channel state information is challenging, there are many effective methods proposed in existing work.

Through the multi-hop RIS communication, the foreign car $A$ would simultaneously receive multiple propagated models of local flying cars in a large scope, even in environments with significant signal blockage.

\subsection{KD-based Model Integration}
Car $A$ must integrate the knowledge from the multiple propagated DL models it receives.
Traditional DL model integration methods, such as federated learning, typically average the parameters of received DL models, assuming identical structures \cite{10329973Zhou}. However, this approach fails when the received DL models have diverse structures.

Unfortunately, due to varying onboard computing and battery capacities, flying cars may generate propagated DL models with different structures. To manage this diversity, we propose a KD-based integration method for non-identical DL model integration.

\begin{figure}[h]
\centering
     \includegraphics[width=.48\textwidth]{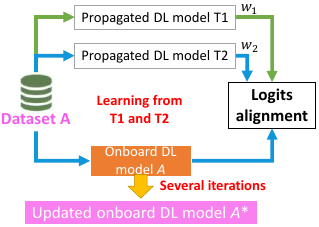} 
     \caption{Non-identical Model Integration.} 
\label{fusion_map}
\end{figure}

Fig.~\ref{fusion_map} illustrates the process of integrating non-identical DL models. Resorting to the extensive coverage of the multi-hop RIS communication, car $A$ can receive multiple propagated DL models from different local cars simultaneously. 
As described, car A receives the propagated models from cars $T1$ and $T2$, respectively, which function as teachers in a multi-teacher-one-student knowledge distillation (KD) framework. The primary objective is to distill the knowledge from the propagated model $T1$ and $T2$ into the onboard DL model of car $A$.

We have further clarified this process. Specifically, we sample point cloud data from dataset $A$ and feed it separately into models $T1$ and $T2$. This produces the predicted vectors $Z^{R_{T1}}$ and $Z^{R_{T2}}$, respectively. The onboard model $A$ is then trained using a combination of the true point cloud classification label $\bm{y}_i$, the prediction vector $Z^{R_{T1}}$ from model $T1$, and the prediction vector $Z^{R_{T2}}$ from model $T2$. The combination KD training objective function is defined as follows:
}

\begin{equation}
\begin{split}
\begin{aligned}
min_{Z} \ L(Z) =\sum_{i=1}^n (1-w_1-w_2)\mathscr{L}(\sigma(\bm{Z^{r}}(\bm{x}_i)), \bm{y}_i) \\
+ w_1\tau^2 \mathscr{L}(\sigma(Z^{R_{T1}}(\bm{x}_i),\tau), \sigma(Z^{r}(\bm{x}_i), \tau)) \\
+ w_2\tau^2 \mathscr{L}(\sigma(Z^{R_{T2}}(\bm{x}_i),\tau), \sigma(Z^{r}(\bm{x}_i), \tau)),
\end{aligned}
\end{split}
\label{dsfjsdhkgeriughprevnkl}
\end{equation}

{
\noindent in which $Z^{R_{T1}}$ and $Z^{R_{T2}}$ are the logits of propagated DL model $T1$ and $T2$, respectively, and $w_1$, $w_2$ are their alignment weights.

where the weight $w_1$ reflects the influence of $Z^{R_{T1}}$ on the training of onboard model $A$, while $w_2$ represents the influence of $Z^{R_{T2}}$. Additionally, $(1-w_1-w_2)$ captures the impact of the true point cloud label $\bm{y}_i$ on model $A$'s training.

As demonstrated, for every data sample from dataset $A$, the classifications from $Z^{R_{T1}}$, $Z^{R_{T2}}$, and the true label $\bm{y}_i$ are utilized to train the parameters of model $A$.
Through this process (After several iterations), the onboard DL model of car $A$ attains the environmental knowledge from the propagated DL models of $T1$ and $T2$, evolving into an updated DL model, $A^*$.
This process facilitates the inheritance and integration of knowledge from the well-trained local DL models.

Importantly, the KD process does not impose any structural requirements on models $T1$ and $T2$. Similarly, the onboard model A does not need to match the model structure of $T1$ or $T2$. This flexibility allows for seamless integration of non-identical DL models through the KD process, thus enabling efficient knowledge transfer across distinct models.
Note that Car $A$ can also select a type of the onboard model structure as the student in the KD-based integration, maximizing the efficiency of the knowledge integration.

}


\section{System Optimization}

This section introduces the proposed optimization problem and the related policy used to obtain the optimal DL model compression, propagation, and integration in multi-hop RIS-aided DL model sharing.

We consider a scenario that involves $I_m$ types of propagated DL models, $1, \ldots, I_m$, for car $Tm$, and $I_A$ types of onboard DL models, $1, \ldots, I_A$, for car $A$. Without loss of generality, we assume that the storage size of DL models increases with their indices for both cars $T_m$ and $A$.

Large DL models may perform worse than lightweight models due to overfitting. However, onboard DL models tend to overfit less due to the limited computing power and battery capacity of vehicles. To simplify the analysis, we assume that larger DL models generally have better generalization ability and perform better.

To avoid potential accidents, flying car $A$ must detect air entities and ground pedestrians accurately within a tolerable time, $\mathcal{T}_{\max}$. This time constraint is expressed as:
\begin{equation}
    \max_m \left\{\frac{S_{m}(i_m)}{R_m} \right\} + \mathcal{T}^\mathrm{KD}_A(i_A) \leq \mathcal{T}_{\max},
\end{equation}
\noindent where $i_m$ and $i_A$ represent the indices of the DL models chosen by cars $T_m$ and $A$, respectively. $S_{m}(i_m)$ is the size of the $i_m$-th DL model for car $T_m$, and $\mathcal{T}^\mathrm{KD}_A(i_A)$ is the estimated KD integration time for the $i_A$-th DL model on car $A$.
The KD integration time, $\mathcal{T}^\mathrm{KD}_A(i_A)$, is defined as the time required for car $A$ to integrate the propagated DL model and achieve an updated accuracy that exceeds an acceptable threshold. It is calculated as:
\begin{equation}
    \mathcal{T}^\mathrm{KD}_A(i_A) = \frac{F(i_A)}{f_A},
\end{equation}
\noindent where $F(i_A)$ represents the floating-point operations required during KD, and $f_A$ is the computing frequency of the chips on car $A$. $\mathcal{T}^\mathrm{KD}_A(i_A)$ depends on the DL models selected for integration by $T_1$, $T_2$, and $A$.
Thus, the overall optimization problem can be formulated as:
\begin{IEEEeqnarray}{lCl}
    \text{\theprob\label{prob.origin}: } 
    &\max_{\mathbf{\Theta}, \mathcal{I}} \ &\sum_{m=1}^{M} w_{m} i_m + w_{M+1} i_{A} \nonumber\\
    && - w_{M+2} \max_m \left\{\frac{S_{m}(i_m)}{R_m} \right\} \IEEEyesnumber \IEEEyessubnumber \label{equ.optimize_obj}\\
    &\text{s.t.} &i_m \in \{1, \ldots, I_m\}, \quad \forall m \IEEEyessubnumber \\
    && i_A \in \{1, \ldots, I_A\}, \IEEEyessubnumber \\
    &&0 \leq \theta_{i, j} \leq 2\pi, \quad \forall i, j \IEEEyessubnumber \label{equ.phase_constraint}\\
    && \max_m \left\{\frac{S_{m}(i_m)}{R_m} \right\} + \mathcal{T}^\mathrm{KD}_A(i_A) \leq \mathcal{T}_{\max}, \IEEEyessubnumber
\end{IEEEeqnarray}
where $w_{1}, w_{2}, \dots, w_{M+2}$ are weight parameters, $\mathbf{\Theta} = \{\mathbf{\Theta}_i| \forall i\}$ and $\mathcal{I} = \{i_m|\forall m\} \cup \{i_A\}$.
The objective of \probref{prob.origin} consists of two parts: the first part, $\sum_{m=1}^{M} w_{m} i_m + w_{M+1} i_A$, maximizes the indices of the DL models selected by $T_m$ and $A$, where a higher index corresponds to a larger DL model with better detection accuracy. The second part, $- w_{M+2} \max_m \left\{\frac{S_{m}(i_m)}{R_m} \right\}$, minimizes the maximum transmission delay of the propagated DL models from the $M$ local cars.

{
In our optimization model P1, the trade-off between performance loss and storage size reduction is managed by adjusting the weight parameters $(w_{1}, w_{2}, \dots, w_{M+2})$. Here, $(w_{1}, w_{2}, \dots, w_{M+1})$ reflects the system’s emphasis on DL performance, while $w_{M+2}$ represents the system’s focus on transmission delay.

When the weights $(w_{1}, w_{2}, \dots, w_{M+1})$ are increased, DL performance is prioritized, leading to the selection of a KD model with a larger storage size. Conversely, when the transmission weight $w_{M+2}$ is increased, the focus shifts towards reducing the storage size of the KD model. Thus, by adjusting the weight parameters $(w_{1}, w_{2}, \dots, w_{M+2})$ in P1, we can effectively balance model performance loss against storage size reduction.

\subsection{Multi-hop RIS Optimization}
The optimization problem \probref{prob.origin} is challenging to solve directly due to its mixed-integer nonlinear nature. Therefore, we decompose \probref{prob.origin} into two simpler sub-problems: multi-hop RIS optimization and KD-based DL model selection.
First, we fix the selected DL models $\mathcal{I}$ and focus on optimizing the multi-hop RIS configuration:
\begin{IEEEeqnarray}{lCl}
    \text{\theprob\label{prob.com_subprob}: }& \min_{\mathbf{\Theta}}\ & \max_m \left\{\frac{S_{m}(i_m)}{R_m}\right\} \IEEEyesnumber \IEEEyessubnumber \\
    & \text{s.t. } &0 \leq \theta_{i,j} \leq 2\pi, \quad \forall i, j \IEEEyessubnumber \\
    && \max_m \left\{\frac{S_{m}(i_m)}{R_m}\right\} + \mathcal{T}^\mathrm{KD}_A(i_A) \leq \mathcal{T}_{\max}. \IEEEyessubnumber \label{equ.comm_constraint_opt}
\end{IEEEeqnarray}

\noindent However, solving \probref{prob.com_subprob} is still difficult due to the non-convex objective and constraint \eqref{equ.comm_constraint_opt}. To address this, we substitute \eqref{sdafkhjaguyfgercbbaaosduf} into \eqref{equ.comm_constraint_opt} and transform it as:
\begin{equation}
\begin{aligned}
    & \max_m \bigg\{ \frac{S_{m}(i_m)}{R_m}\bigg\} \leq \mathcal{T}_{\max} - \mathcal{T}^\mathrm{KD}_A(i_A) \\
    & \Leftrightarrow \min_m \zeta_m\left|\left(\prod_{i=1}^{N} \mathbf{\Theta}_{i} \mathbf{H}_{i}\right) \boldsymbol{g}_{m} \right|^2 \geq 1,
\end{aligned}
\label{dasfgaeruyfguskjlk}
\end{equation}

\noindent where,
\begin{equation}
    \zeta_m = \frac{\left| x_m \right|^2}{\sigma_A^2\left( 2^{\frac{S_{m}(i_m)}{B_m\left(\mathcal{T}_{\max} - \mathcal{T}^\mathrm{KD}_A(i_A)\right)}} - 1 \right)}.
\end{equation}

\noindent To simplify, maximizing the channel gain is equivalent to minimizing transmission delay. Thus, \probref{prob.com_subprob} becomes:
\begin{IEEEeqnarray}{lCl}
    \text{\theprob\label{prob.com_tf1}: }& \max_{\mathbf{\Theta}}\ & \min_m \ \zeta_m\left|\left(\prod_{i=1}^{N} \mathbf{\Theta}_{i} \mathbf{H}_{i}\right) \boldsymbol{g}_{m} \right|^2 \IEEEyesnumber \IEEEyessubnumber \\
    & \text{s.t. } &0 \leq \theta_{i,j} \leq 2\pi, \quad \forall i, j \IEEEyessubnumber \\
    && \min_m \ \zeta_m\left|\left(\prod_{i=1}^{N} \mathbf{\Theta}_{i} \mathbf{H}_{i}\right) \boldsymbol{g}_{m} \right|^2 \geq 1 \IEEEyessubnumber
\end{IEEEeqnarray}

We apply the Block Coordinate Descent (BCD) method \cite{9420314Ikeshita} to optimize \probref{prob.com_tf1} iteratively. \probref{prob.com_tf1} can be decomposed into $M$ sub-problems, each focusing on optimizing one-hop RIS phase shifts. Specifically, we alternately optimize the phase shifts of each RIS while keeping the others fixed.
The optimization of a single RIS can be formulated as:
\begin{IEEEeqnarray}{lCl}
    \text{\theprob\label{prob.com_tf2}: }& \max_{\mathbf{\Theta}_n}\ & \min_m \ \zeta_m\left|\boldsymbol{h}_n \mathbf{\Theta}_{n} \boldsymbol{g}_{m,n} \right|^2 \IEEEyesnumber \IEEEyessubnumber \\
    & \text{s.t. } &0 \leq \theta_{n,j} \leq 2\pi, \quad \forall j \IEEEyessubnumber
\end{IEEEeqnarray}
where $\boldsymbol{h}_n = \prod_{i=n+1}^{N} \mathbf{\Theta}_{i} \mathbf{H}_{i}$, $\boldsymbol{g}_{m,n} = \mathbf{H}_{n} (\prod_{i=1}^{n-1} \mathbf{\Theta}_{i} \mathbf{H}_{i}) \boldsymbol{g}_{m}$.

By defining $v_{n,i} = e^{j \theta_{n,i}}$, $\boldsymbol{v}_n = [v_{n,1}, \ldots, v_{n,N_n}]^H$, and $\mathbf{\Phi}_{m,n} = \mathrm{diag}(\boldsymbol{h}_n) \boldsymbol{g}_{m,n}$, \probref{prob.com_tf2} can be rewritten as a standard form of Quadratically Constrained Quadratic Program (QCQP) problem:
\begin{IEEEeqnarray}{lCl}
    \text{\theprob\label{prob.com_tf3}: }& \max_{\mathbf{\Theta}_n}\ & \min_m \ \zeta_m\boldsymbol{v}_n^H \tilde{\mathbf{\Phi}}_{m,n} \boldsymbol{v}_n \IEEEyesnumber \IEEEyessubnumber \\
    & \text{s.t. } &|v_{n,j}|^2=1, \quad \forall j \IEEEyessubnumber
\end{IEEEeqnarray}
where $\tilde{\mathbf{\Phi}}_{m,n} = \mathbf{\Phi}_{m,n} \mathbf{\Phi}_{m,n}^H$.
Define $\mathbf{V}_{n} = \boldsymbol{v}_n \boldsymbol{v}_n^H$, the above \probref{prob.com_tf3} can be solved by the semideﬁnite relaxation (SDR) method. 
SDR is widely recognized as an efficient and high-performance method, often used to solve semidefinite programming problems with polynomial complexity \cite{5447068Luo}. According to the standard form of SDR, \probref{prob.com_tf3} is then revised into:
\begin{IEEEeqnarray}{lCl}
        \text{\theprob\label{prob.com_tf4}: }& \max_{\mathbf{V}_{n}}\ & \min_m \ \zeta_m\mathrm{tr}(\tilde{\mathbf{\Phi}}_{m,n} \mathbf{V}_{n}) \IEEEyesnumber \IEEEyessubnumber \\
    & \text{s.t. } & v_{n, i, i} = 1, \quad \forall i \IEEEyessubnumber \\
    && \mathbf{V}_{n} \succ 0 \IEEEyessubnumber
\end{IEEEeqnarray}
where $v_{n,i,i}=1$ is the $i$th diagonal element of $\mathbf{V}_n$. 

As described in \probref{prob.com_tf4}, the objective function aims to maximize the minimum weighted RIS channel gain $\mathrm{tr}(\tilde{\mathbf{\Phi}}_{m,n} \mathbf{V}_{n})$. 
The weight parameter $\zeta_m$ reflects the communication Quality of Service (QoS) needs for the $m$-th vehicle. Notably, a vehicle with higher communication demands is assigned a greater weight $\zeta_m$. By adjusting these RIS communication weights $\zeta_m$, we can effectively balance the communication needs of multiple vehicles simultaneously.

This problem can be solved using convex optimization tools like CVX in Matlab.
It is observed that $\mathrm{rank}(\mathbf{V}_{n}) = \mathrm{rank}(\boldsymbol{v}_n \boldsymbol{v}_n^H) = 1$.
However, the solution of \probref{prob.com_tf4} may not always equivalent to that of \probref{prob.com_tf3} because the SDR relaxation may not always yield a rank-1 solution.
Therefore, additional operations are needed to extract a valid $\mathbf{V}_n$ and $\boldsymbol{v}_n$.

First, we perform the eigenvalue decomposition operation to $\mathbf{V}_n$, i.e., $\mathbf{V} = \mathbf{U}\Sigma \mathbf{U}^H$, in which $\mathbf{U}$ is a unitary matrix and $\Sigma$ is a diagonal matrix.
Subsequently, a sub-optimal solution of \probref{prob.com_tf3} is given as $\boldsymbol{v}_n = \mathbf{U}\Sigma^{\frac{1}{2}}\boldsymbol{r}$, where $\boldsymbol{r} \in \mathbb{C}^{N_n}$ is a random vector following the circularly symmetric complex Gaussian distribution with zero mean and covariance matrix $\mathbf{I}_{N_n.}$.
Finally, the solution $\boldsymbol{v}_n$ of \probref{prob.com_tf3} is obtained by the randomly generated $\boldsymbol{r}$ to achieve the optimal objective value.

In the worst-case scenario, the computational complexity of SDR is $\mathcal{O}\left(\max \{m, n\}^4 n^{1 / 2} \log (1 / \epsilon)\right)$, where $m$ represents the number of constraints, $n$ is the problem size, and $\epsilon$ denotes the solution accuracy \cite{5447068Luo}. However, due to the involvement of the Gaussian randomization process, SDR requires an additional process to generate variables in order to obtain a valid solution, which incurs an extra computational complexity of $\mathcal{O}(LN_i^3)$, where $L$ represents the number of iterations.

\subsubsection{\textbf{Fast RIS Phase Shift Optimization}}
In practical applications, reducing the total time consumption of the optimization process is often more critical than merely improving transmission rates. Therefore, we propose a fast phase shift optimization method for multi-hop RIS implementation in Urban Air Mobility (UAM).

Given $\boldsymbol{a}_{m,n} = \mathrm{diag}(\boldsymbol{h}_n)\boldsymbol{g}_{m,n}$, we have:
\begin{equation}
      \boldsymbol{h}_n \mathbf{\Theta}_{n} \boldsymbol{g}_{m,n} = \boldsymbol{v}_n^H \boldsymbol{a}_{m,n}.
\end{equation}
Notably, the expression $\boldsymbol{v}_n^H \boldsymbol{a}_{m,n}$ can be viewed as a sum of multiple complex scalars.
It is well known that $|\boldsymbol{v}_n^H \boldsymbol{a}_{m,n}|$ is maximized when the phases of all vectors are identical. The optimal $\boldsymbol{v}_n$ is thus:
\begin{equation}
    \label{equ.optimal_v_for_tm}
    \boldsymbol{v}_{m,n}^* = e^{j \mathrm{arg}(\boldsymbol{a}_{m,n})}.
\end{equation}

\noindent However, Eq.~\eqref{equ.optimal_v_for_tm} only maximizes the channel gain for car $Tm$.
Since all local cars share the same RIS phase shift $\mathbf{\Theta}_n$, there are insufficient degrees of freedom to fully satisfy this condition for all $M$ transmitters. Therefore, we compute an optimal $\bar{\boldsymbol{v}}_n$ as the sum of the expected phase shifts for all transmitters:
\begin{equation}
    \bar{\boldsymbol{v}}_n = e^{j \mathrm{arg} \left(\sum_{m=1}^M\frac{\boldsymbol{a}_{m,n}}{\zeta_m|\boldsymbol{a}_{m,n}|^{\beta_a}}\right)},
    \label{aksjdfgiuyewgfuysdfgaeuy}
\end{equation}

\noindent where $\beta_a$ is a predetermined parameter. The optimal phase shift of the $n$-th RIS is then given by $\mathrm{arg}(\bar{\boldsymbol{v}}_n)$.

\subsection{KD-based DL Model Selection}
Given the optimal RIS phase shifts $\mathbf{\Theta}$, the KD-based DL model selection problem can be expressed as:
\begin{IEEEeqnarray}{lCl}
\text{\theprob\label{prob.kd_select}: } 
    &\max_{\mathcal{I}} \ &\sum_{m=1}^{M} w_{m} i_m + w_{M+1} i_{A} \IEEEyesnumber \IEEEyessubnumber \\
    &\text{s.t.} &i_m \in \{1, \ldots, I_m\}, \quad \forall m \IEEEyessubnumber \\
    && i_A \in \{1, \ldots, I_A\}, \IEEEyessubnumber \\
    && \max_m \left\{\frac{S_{m}(i_m)}{R_m} \right\} + \mathcal{T}^\mathrm{KD}_A(i_A) \leq \mathcal{T}_{\max} \IEEEyessubnumber
\end{IEEEeqnarray}

This is a nonlinear integer programming problem that can be transformed into a zero-one integer programming by introducing indicator variables $z_{m,i}$ and $z_{A,i}$. Here, $z_{m,i} = 1$ indicates that car $T_m$ selects the $i$-th DL model, otherwise $z_{m, i} = 0$. $z_{A,i}$ is the indicator for car $A$'s DL model selection. The reformulated problem is:
\begin{IEEEeqnarray}{lCl}
    \text{\theprob\label{prob.kd_select2}: } 
    &&\max_{\mathcal{Z}} \ \sum_{m=1}^{M} \sum_{i=1}^{I_m} w_{m} i z_{m, i}  + \sum_{i=1}^{I_A} w_{M+1} i z_{A, i},  \IEEEyesnumber \IEEEyessubnumber \\
    &\text{s.t.} &\sum_{i=1}^{I_m} z_{m, i} = 1, \quad \forall m \IEEEyessubnumber \\
    &&\sum_{i=1}^{I_A} z_{A, i} = 1, \IEEEyessubnumber \\
    &&\sum_{i=1}^{I_m}  \frac{S_{m}(i)z_{m, i}}{R_m} + \sum_{i=1}^{I_A} \mathcal{T}^\mathrm{KD}_A(i)z_{A, i} \leq \mathcal{T}_{\max}, \IEEEyessubnumber
\end{IEEEeqnarray}

\noindent This linear zero-one integer programming problem can be solved using the CVX toolbox in Matlab.
\begin{algorithm}[htb]
    \caption{Multi-hop RIS-aided DL Model Propagation (MRMP)}
    \label{alg.model_optimization}
    \KwInitialize{$\mathcal{I}^{(0)} \leftarrow$ smallest models,
    $\mathbf{\Theta}^{(0)} \leftarrow$ a random matrix}
    \For{$k \leftarrow 0, \ldots, k_{\max}-1$}{
        \For{$n \leftarrow 1, \ldots, N$}{
            Given $\mathcal{I}^{(k)}$, $\{\mathbf{\Theta}_i^{(k+1)}| \forall i < n\}$, and $\{\mathbf{\Theta}_i^{(k)}| \forall i > n\}$, solve \probref{prob.com_tf4}\;
            obtain $\mathbf{\Theta}_i^{(k+1)}$ according to the optimal solution to \probref{prob.com_tf4}\;
        }
        Given $\mathbf{\Theta}^{(k+1)}$, solve \probref{prob.kd_select}\;
        $\mathcal{I}^{(k+1)} \leftarrow$ the optimal solution to \probref{prob.kd_select}\;
    }
    \Return $\mathcal{I}^{(k_{\max})}, \mathbf{\Theta}^{(k_{\max})}$.
\end{algorithm}

Furthermore, an iterative algorithm is proposed to jointly optimize both the RIS phase shift and DL model selection. 
In particular, the algorithm begins by initializing the smallest DL models for both the propagated and onboard models, while the multi-hop RISs are initialized with random phase shift matrices. 
At each iteration, the optimal RIS phase shift matrices are determined by solving \probref{prob.com_tf4}. Once the RIS configurations are fixed, the algorithm selects the most suitable DL model. This alternating optimization process continues until the maximum number of iterations is reached.

The workflow of the Multi-hop RIS-aided DL Model Propagation (MRMP) scheme is depicted in Alg.~\ref{alg.model_optimization}. 
The DL models $\mathcal{I}$ are initialized using the smallest models, while the RIS phase shift matrices $\mathbf{\Theta}$ are set randomly. 
Following this, the proposed Multi-hop RIS-aided DL Model Propagation (MRMP) algorithm applies the BCD method to derive a suboptimal RIS shift matrix. 
Once $\mathbf{\Theta}$ is obtained, the optimal DL model is selected by solving \probref{prob.kd_select}. 
This sequence of steps is repeated until the algorithm converges, typically after $k_{\max}$ iterations.

}

\begin{figure}[h]
\centering
     \includegraphics[width=.43\textwidth]{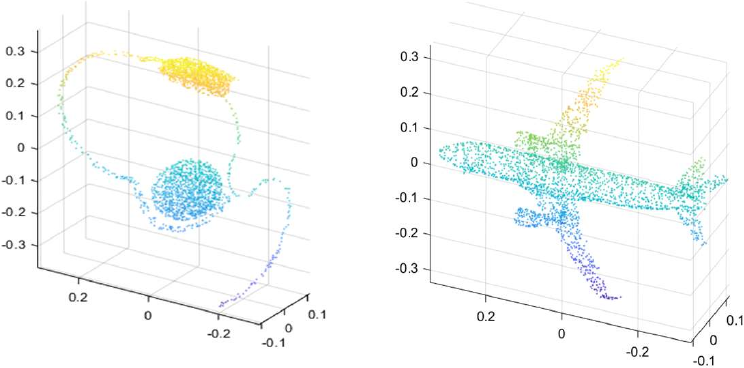} 
     \caption{Earphone and aircraft point cloud data.} 
\label{dataset_demo}
\end{figure}

{
\section{Performance Evaluation}
To assess the effectiveness of the proposed DL model sharing scheme for point cloud learning, we conduct experiments using the popular point cloud dataset, ShapeNet \cite{3326362YueWang}. 
This dataset includes $16681$ point cloud samples categorized into $16$ classes. 
An example of point cloud representations for an earphone and an aircraft is shown in Fig.~\ref{dataset_demo}. 
In our experiments, the training dataset comprises approximately $2\%-8\%$ of the total ShapeNet dataset, while the test dataset covers about $80\%$. 
The proposed algorithms are implemented on a computer equipped with an NVIDIA GeForce RTX 2060 GPU, using PyTorch for model training. 
The machine also features an Intel(R) Core(TM) i7-9700K CPU and 32GB of RAM, with the primary simulation parameters summarized in Tab.~\ref{table_2}.
}

\begin{table}[htb]
\begin{center}   
\caption{Simulation parameters}  
\label{table_2} 
\begin{tabular}{|c|c|c|c|}   
\hline   learning rate & 0.001 & batch size & 32 \\ 
\hline   KD Temp $\tau$ & $[1,15]$ & KD weight $\alpha$ & $[0,0.4]$ \\     
\hline   Storage of \textit{Model 1} & $280K$ & Storage of \textit{Model 2} & $2.3M$ \\    
\hline   Storage of \textit{Model 3}  & $4.1M$ & Storage of \textit{Pointnet} & $6.2M$ \\
\hline Number of RIS elements & 64 & Number of RIS & 3 \\
\hline Racian factor $\beta_H$ & 1 & Bandwith & 240 MHz \\
\hline
\end{tabular}   
\end{center}   
\end{table}

\begin{figure*}
\centering
\subfloat[DL model compression performance.]{\label{KD_performance}{\includegraphics[width=0.33\linewidth]{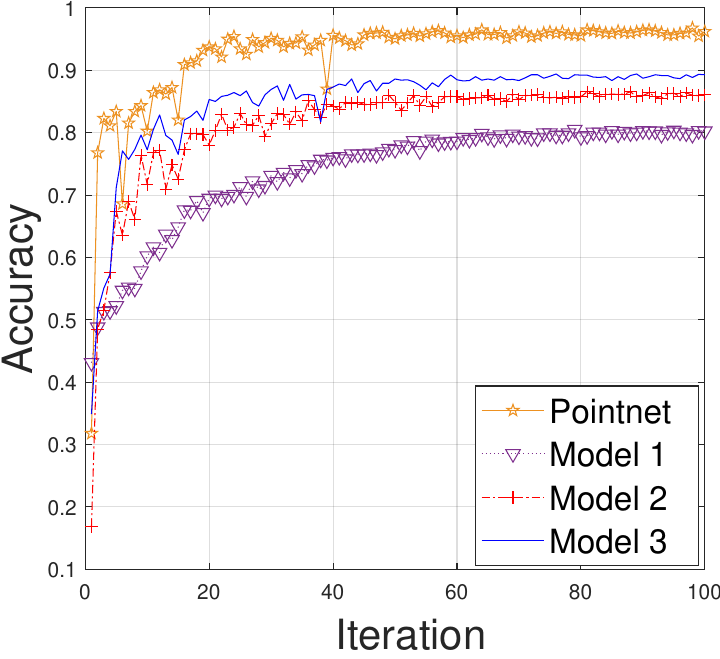}}} \hfill
\subfloat[DL model integration performance.]{\label{model_int_curve}{\includegraphics[width=0.324\linewidth]{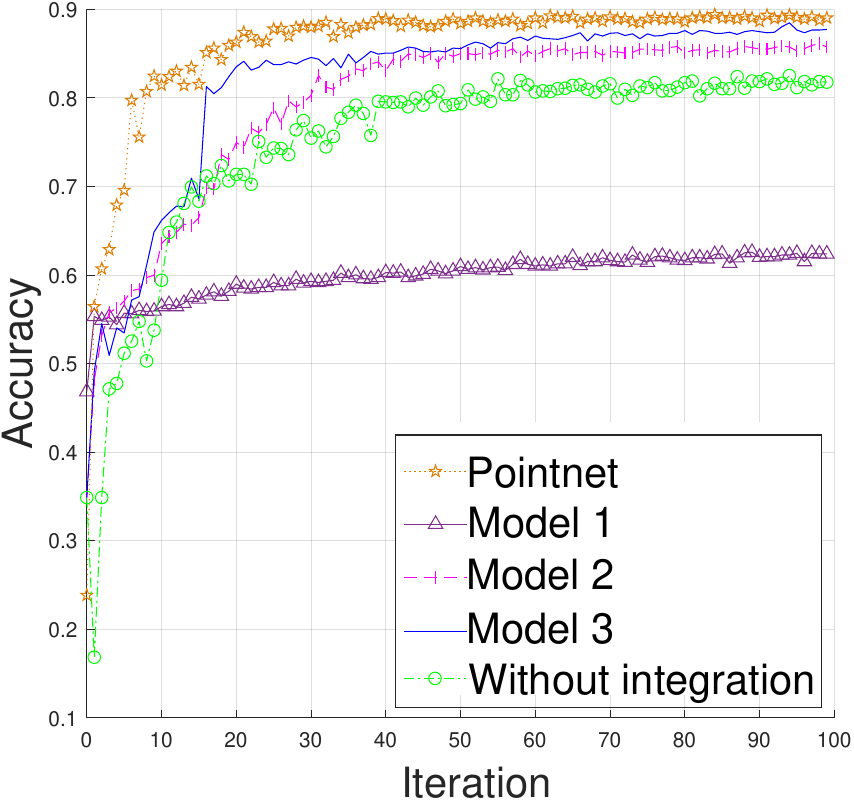}}} \hfill
\subfloat[Different propagated DL model Integration.]{\label{Model_integration_with_diff_tch}{\includegraphics[width=0.323\linewidth]{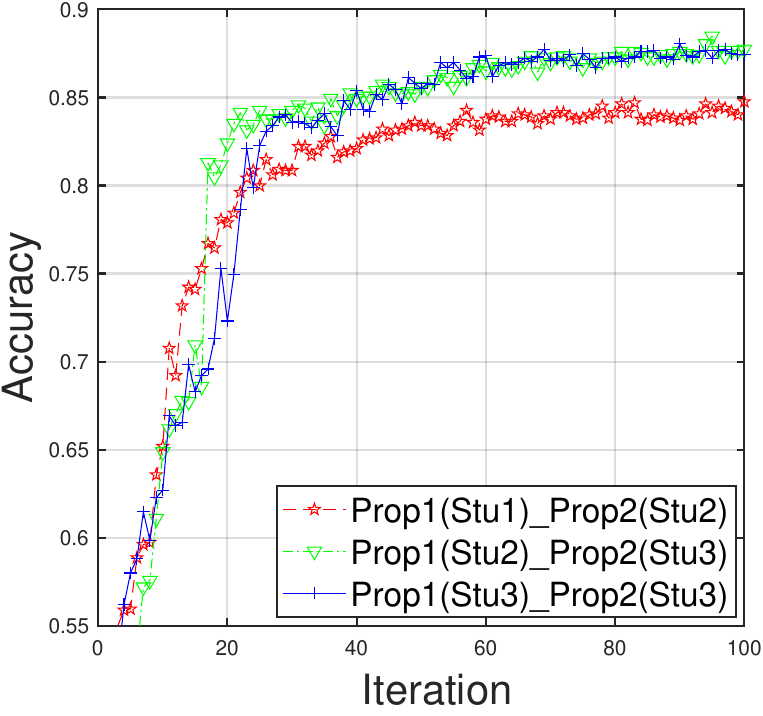}}} \hfill
\caption{KD-based DL model compression and integration performance.}
\label{KD_Compression_integration}
\end{figure*}

\begin{figure}
\centering
\subfloat[DL model 1 with $280K$ storage size]{\label{DL model 1}{\includegraphics[width=0.37\linewidth]{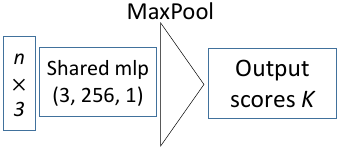}}} \\
\subfloat[DL model 2 with $2.3M$ storage size.]{\label{DL model 2}{\includegraphics[width=0.85\linewidth]{image/student2.pdf}}} \\
\subfloat[DL model 3 with $4.1M$ storage size.]{\label{DL model 3}{\includegraphics[width=0.98\linewidth]{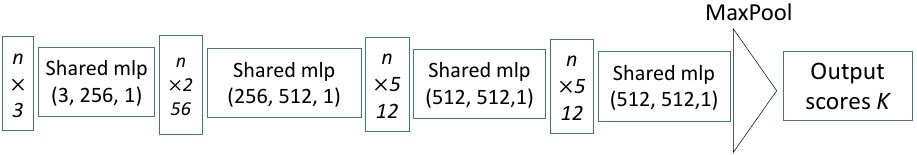}}} \\
\caption{Three proposed DL model structures and their storage sizes.}
\label{ThreeStudentStructures}
\end{figure}

{
\subsection{KD-based Compression and Integration}
In general, due to the nonlinear and complex structure of neural networks, there is no explicit functional relationship between the degree of KD compression and the resulting performance loss. Furthermore, the performance of a DL model is highly task-dependent, meaning that the optimal model structure varies according to task complexity. Typically, as the size and parameters of a deep learning model decrease, its performance also diminishes. The most reliable method for quantifying this trade-off is through direct performance testing. 
}

The detection accuracy testing of the KD-based DL models is illustrated in Fig.~\ref{KD_performance}. 
The $x$-axis represents the number of training iterations, while the $y$-axis indicates the point cloud detection accuracy. 
In this setup, Pointnet serves as the onboard DL model that guides the training of three distinct propagated models. 
The structures of these three propagated models are displayed in Fig.~\ref{ThreeStudentStructures}.

Due to its simpler architecture, DL model 1 exhibits the lowest detection accuracy. 
However, DL models 2 and 3, by leveraging the knowledge distilled from Pointnet, achieve significantly higher detection performance. 
Overall, the KD process effectively reduces the storage size of the DL models, with only a slight performance loss.

Fig.~\ref{model_int_curve} presents the DL model integration performance for various propagated models and the onboard model of car $A$. 
As illustrated in Fig.~\ref{fusion_map}, the integration involves two KD teachers (propagated models from $T1$ and $T2$) and one student (the onboard model of car $A$). 
We assign DL model 2 to $T1$ and DL model 3 to $T2$ to evaluate the integration performance. 
Four different DL models are selected as the onboard model of car $A$ to distill knowledge from these propagated models: DL model 1, DL model 2, DL model 3, and Pointnet. 
As expected, the detection accuracy improves with the DL model size. 
For example, Fig.~\ref{model_int_curve} compares the performance of DL model 3 with and without integration, using $2\%$ of the ShapeNet dataset for training and $80\%$ for testing. 
Without integration, DL model 3 achieves an accuracy of $81\%$, whereas integrating the propagated models increases the accuracy to $87\%$, yielding a $6\%$ improvement for car $A$.

\begin{figure}
\centering
     \includegraphics[width=.38\textwidth]{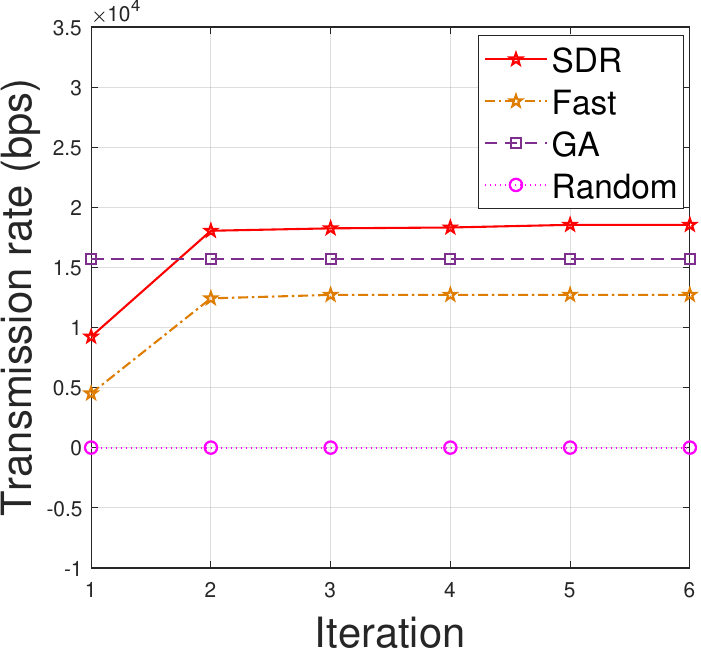} 
     \caption{Transmission rate of different multi-hop RIS schemes.} 
\label{multi_RIS_performance}
\end{figure}

Fig.~\ref{Model_integration_with_diff_tch} compares the integration performance of different propagated DL models shown in Fig.~\ref{fusion_map}. 
This simulation examines various combinations of two propagated models, including DL model 1 paired with model 2, DL model 2 paired with model 3, and DL model 3 pair. 
In Fig.~\ref{Model_integration_with_diff_tch}, $Prop1(Stu1)\_Prop2(Stu2)$ represents the combination of DL model 1 as $T1$ and DL model 2 as $T2$. 
Similarly, $Prop1(Stu2)\_Prop2(Stu3)$ and $Prop1(Stu3)\_Prop2(Stu3)$ represent other combinations. 
Interestingly, both $Prop1(Stu2)\_Prop2(Stu3)$ and $Prop1(Stu3)\_Prop2(Stu3)$ achieve similar detection accuracy ($87\%$), while $Prop1(Stu1)\_Prop2(Stu2)$ performs slightly worse at $84\%$. 
This is because DL model 3 has the best performance among the three models, and its presence in both $Prop1(Stu2)\_Prop2(Stu3)$ and $Prop1(Stu3)\_Prop2(Stu3)$ elevates their detection accuracy. 
On the other hand, $Prop1(Stu1)\_Prop2(Stu2)$ is limited by the lower performance of DL model 2, highlighting that the integration performance is largely influenced by the more capable propagated model.

{
\subsection{Transmission Rate Comparisons}
Fig.~\ref{multi_RIS_performance} illustrates the transmission rates achieved by four multi-hop RIS optimization algorithms. 
The SDR, our proposed SDR-based BCD optimization algorithm, achieves the highest transmission rate. 
The second-best performance is delivered by the Genetic Algorithm (GA), which differs from BCD in that it employs a heuristic, global search approach for optimizing RIS phase shifts. 
Unlike the iterative nature of the SDR and GA algorithms, the Random algorithm selects RIS phase shifts arbitrarily, which is why its performance appears as a flat line in the figure. 
Additionally, the configuration for the three-multi-hop-RIS transmission is shown in Fig.~\ref{simulation_scenario}.
\begin{figure}
\centering
     \includegraphics[width=.48\textwidth]{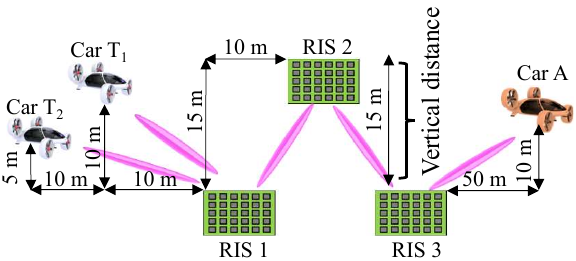} 
     \caption{Configurations of three-multi-hop-RIS transmission.} 
\label{simulation_scenario}
\end{figure}

As expected, the SDR algorithm converges to the highest transmission rate since it provides an approximately optimal solution. 
The GA algorithm, although less efficient, also performs well and converges to a high transmission rate. 
Since the iteration of the GA algorithm is quite different from the BCD iteration, we depict a horizontal line to reflect the final converged transmission rate of the GA algorithm.
In contrast, the Random algorithm yields much lower transmission rates because it cannot ensure proper alignment of the multiple RISs, which is critical for achieving high transmission rates in a multi-hop setting.

\begin{figure}
\centering
     \includegraphics[width=.38\textwidth]{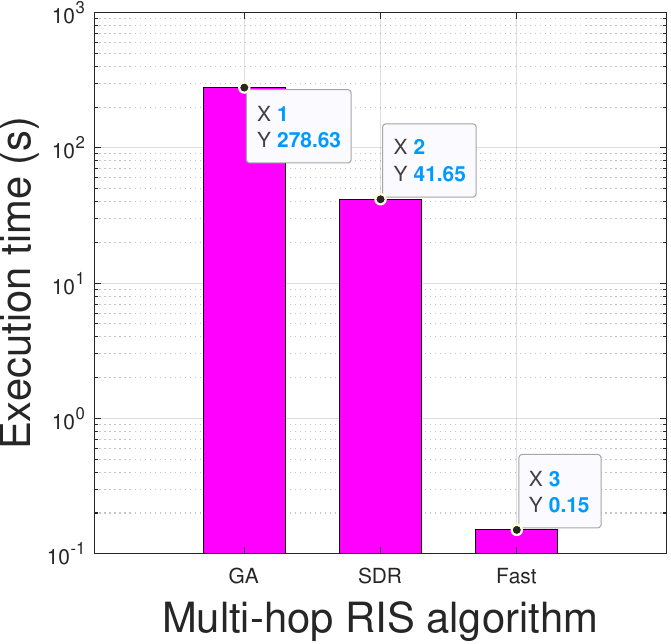} 
     \caption{Execution time of multi-hop RIS optimization algorithms.} 
\label{time_complexity}
\end{figure}

Fig.~\ref{time_complexity} compares the execution times of the different algorithms. 
Due to the extensive crossovers and selections in the GA process, it takes around $278$ seconds to complete. 
The SDR algorithm, though faster, still incurs significant computation time due to the Gaussian randomization of $v_n$. 
In contrast, the Fast RIS phase shift algorithm, which provides a closed-form solution for the suboptimal phase shift (Eq.~(\ref{aksjdfgiuyewgfuysdfgaeuy})), is much quicker, taking only $0.15$ seconds. 
Although the Fast algorithm does not converge to as high a transmission rate as the SDR, its extremely low execution time makes it highly practical for real-world UAM applications.

Typically, the SDR, GA, and Fast algorithms outperform the Random algorithm in terms of transmission rate. 
This is because multi-hop RIS transmission relies on the sequential alignment of the multiple RISs, which the Random algorithm fails to ensure, resulting in low transmission rates.

\begin{figure}
\centering
     \includegraphics[width=.38\textwidth]{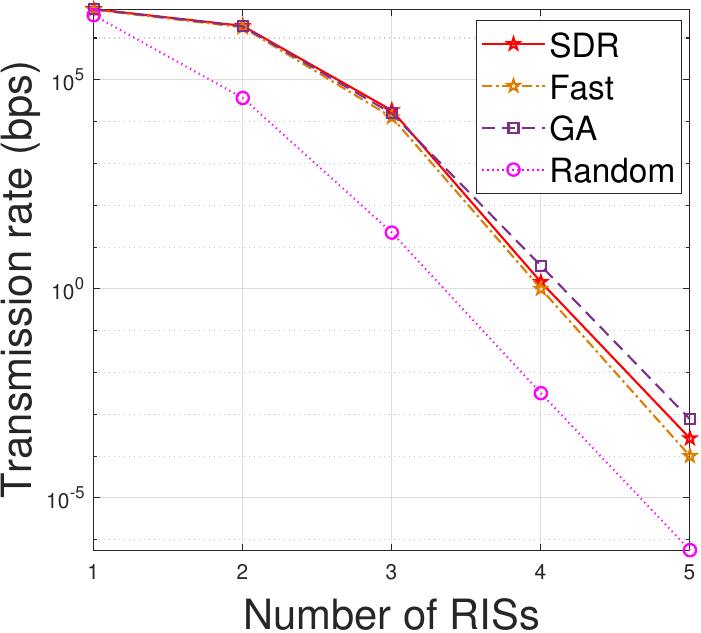} 
     \caption{Transmission rate with the number of RISs.} 
\label{Number_of_RISs}
\end{figure}
Fig.~\ref{Number_of_RISs} shows how the transmission rate changes with the number of multi-hop RISs. 
Interestingly, as the number of RIS reflections decreases, the transmission rate improves. 
This trend is consistent across the SDR, GA, and Fast algorithms. 
When the number of RIS reflections exceeds four, the transmission rate for all algorithms approaches zero.
This phenomenon occurs because RIS reflections and path loss degrade signal power, and since the transmission rate is directly proportional to signal power, multiple hops and longer distances result in reduced transmission rates.

The relative position between successive RISs also influences the transmission rate, as shown in Fig.~\ref{location_scenario}. 
In this experiment, we place RISs at four different azimuth angles with a fixed radius of $30m$. 
Among these, the $30^\circ$ azimuth angle yields the highest transmission rate, as the horizontal component of the reflected signal is maximized when the receiver car $A$ is directly ahead. 
As the azimuth angle increases, the horizontal component diminishes, leading to greater signal dissipation at the receiver.

\begin{figure}
\centering
     \includegraphics[width=.25\textwidth]{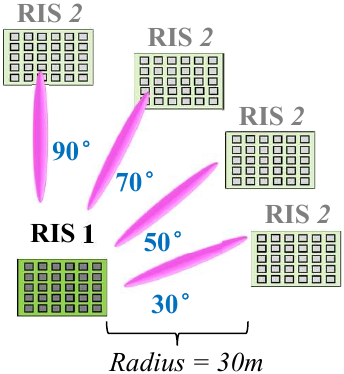} 
     \caption{Demonstration of RIS positions with different azimuth angles.} 
\label{location_scenario}
\end{figure}

\begin{figure}
\centering     \includegraphics[width=.38\textwidth]{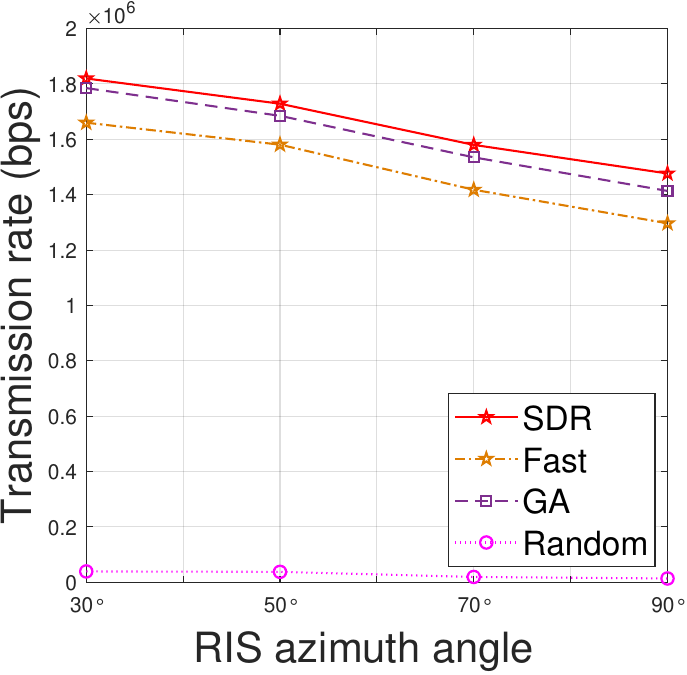} 
     \caption{Transmission rate with different RIS positions.} 
\label{location_degree}
\end{figure}

Fig.~\ref{Vertical_distance_between_RIS} explores the impact of vertical distance between adjacent RISs on the transmission rates for different algorithms. 
The vertical distance setup is illustrated in Fig.~\ref{simulation_scenario}. 
We evaluate the transmission rate for four vertical distances: $15m$, $20m$, $25m$, and $30m$. 
At each distance, the SDR algorithm consistently delivers the highest transmission rate, outperforming the other algorithms. 
As expected, the transmission rates for all algorithms decrease as the vertical distance increases.

\begin{figure}
\centering
     \includegraphics[width=.40\textwidth]{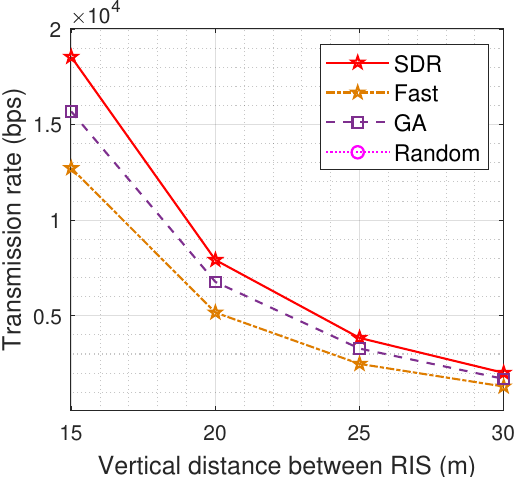} 
     \caption{Transmission rate with different RIS vertical distance.} 
\label{Vertical_distance_between_RIS}
\end{figure}

To further validate the proposed scheme, we simulate its performance with varying numbers of RIS elements, as shown in Fig.~\ref{Number_of_RIS_elements}. 
We consider configurations with $6\times 6$, $7\times 7$, $8\times 8$, and $9\times 9$ RIS elements. 
The results show that as the number of elements increases, the transmission rate improves significantly. 
In fact, when the number of RIS elements increases from $64$ to $81$, the transmission rate experiences a substantial $400\%$ increase.

This improvement arises since RIS phase shift control is typically discrete, limited by the practical manufacturing process. Therefore, the discrete phase shift optimization of RIS can only achieve a suboptimal transmission rate for the system. Fortunately, as the number of RIS elements grows, the degrees of freedom in phase shift control increase, bringing the system closer to optimal transmission rate. Additionally, with more elements, the RIS can capture and reflect more signal energy, enhancing the reflected signal strength.

However, it is important to note that the cost of manufacturing RIS also increases with the number of elements. The production process for large-scale RIS elements is complex, which may reduce yield rates. In practical deployments, a balance must be struck between manufacturing costs and transmission performance gains.

\begin{figure}
\centering
     \includegraphics[width=.38\textwidth]{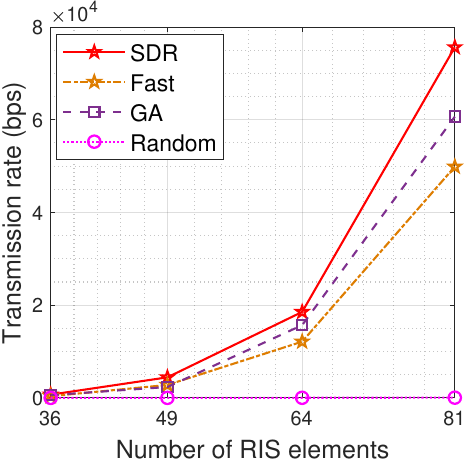} 
     \caption{Transmission rate with different number of RIS elements.} 
\label{Number_of_RIS_elements}
\end{figure}

\begin{figure}
\centering  \includegraphics[width=.37\textwidth]{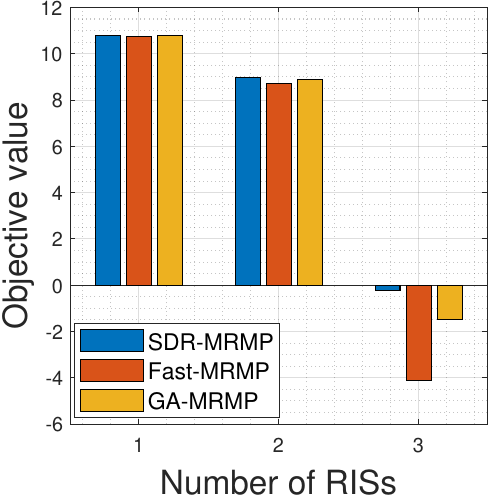} 
     \caption{Objective value of $P1$ with different number of RISs.} 
\label{total_performance_with_number_of_RIS}
\end{figure}

\subsection{System Reward Comparisons}

We also compare the objective value of \probref{prob.origin} across different MRMP algorithms in Fig.~\ref{total_performance_with_number_of_RIS}. 
The SDR-MRMP scheme, which uses SDR to solve \probref{prob.com_tf4} in Alg.~\ref{alg.model_optimization}, achieves the highest objective value. 
The GA-MRMP and Fast-MRMP schemes, which leverage GA and the Fast algorithm, respectively, also perform well, though not as optimally as SDR-MRMP. 
Due to the near-zero transmission rate of the Random algorithm, its results are not included in this figure.
When only a single RIS is used, all MRMP schemes perform similarly. 
However, as the number of RISs increases, the advantage of SDR-MRMP becomes more pronounced. 
Notably, SDR-MRMP improves the total reward by $85\%$ compared to GA-MRMP.

Finally, Tab.~\ref{tab.model_selection} lists the optimal DL model selection for \probref{prob.origin} under different numbers of RISs. 
When only one RIS is employed, all algorithms tend to transmit the onboard model directly, as the channel overhead allows for uncompressed transmission, i.e., the propagated DL model is identical to the onboard model.
However, as more obstacles and RIS hops are introduced, the transmission rate deteriorates, prompting the selection of KD-compressed models to reduce communication overheads. 
In the three-hop RIS setting, both $T1$ and $T2$ select DL model 1, prioritizing communication efficiency over integration performance.
Thus, we can apply different weight $w$ configurations to balance the integration performance and propagation overheads.

\begin{table}[htb]
\centering
\caption{Model selection result}
\label{tab.model_selection}
\begin{tabular}{lcccc}
\toprule
Method    & Number of RIS & Car T1 & Car T2 & Car A \\
\midrule
SDR-MRMP  & 1             & Ponitnet      & Ponitnet      & Ponitnet     \\
GA-MRMP   & 1             & Ponitnet      & Ponitnet      & Ponitnet     \\
Fast-MRMP & 1             & Ponitnet      & Ponitnet      & Ponitnet     \\
SDR-MRMP  & 2             & Ponitnet      & Ponitnet      & Ponitnet     \\
GA-MRMP   & 2             & Model 3      & Model 3      & Ponitnet     \\
Fast-MRMP & 2             & Ponitnet      & Ponitnet      & Ponitnet     \\
SDR-MRMP  & 3             & Model 1      & Model 1      & Ponitnet     \\
GA-MRMP   & 3             & Model 1      & Model 1      & Ponitnet     \\
Fast-MRMP & 3             & Model 1      & Model 1      & Model 3    \\
\bottomrule
\end{tabular}
\end{table}

\section{Conclusion}

In this paper, we presented a multi-hop RIS-aided DL model sharing scheme designed for 3D point cloud detection in Urban Air Mobility (UAM). 
Our approach enhances the onboard detection accuracy of flying cars as they navigate unfamiliar environments. 
To facilitate DL model sharing, we proposed a practical framework that combines multi-hop RIS phase shift control with non-identical model integration, improving propagation throughput, communication range, and knowledge-sharing between local and foreign cars. 
The KD-based DL model compression technique significantly reduces transmission overheads with minimal sharing performance loss. 
Additionally, we developed an approximately optimal SDR algorithm and a fast phase shift optimization strategy to meet the diverse requirements of UAM in terms of transmission rate and optimization time. 
We also devised a novel non-identical DL model integration strategy, enabling knowledge transfer from distinct propagated models. 
The extensive experiments demonstrate that our proposed scheme is both practical and superior to existing benchmarks on popular point cloud datasets. 
This work paves the way for non-identical learning collaboration, showing great potential for large-scale swarm intelligence in heterogeneous systems.
In future work, we will analyze the optimal deployment of RIS in networks and explore stacking intelligent meta surfaces (SIMs) to replace RIS functionality for more efficient aerial multi-hop transmission.

}

%
%



\appendices
\footnotesize
\bibliography{biblio}
\end{document}